\documentclass[journal]{IEEEtran}
\usepackage{ifpdf}
\usepackage[noadjust]{cite}
\usepackage[cmex10]{amsmath}
\usepackage{array}
\usepackage{dblfloatfix}
\usepackage{amssymb}
\usepackage{graphicx}
\usepackage{breqn}
\usepackage{amsthm}
\usepackage{amsmath}
\usepackage{color}
\usepackage{makecell}

\newtheorem{definition}{Definition}[section]
\newtheorem{theorem}{Theorem}

\DeclareMathOperator{\tr}{tr}

\DeclareMathOperator{\diag}{diag}
\DeclareMathOperator{\rank}{rank}
\ifCLASSINFOpdf

\else
\fi
\begin{document}
\large
%
\title{
Robust Radar Detection of a Mismatched Steering Vector Embedded in Compound Gaussian Clutter
}
\author{Mai P. T. Nguyen and
        Iickho Song,~\IEEEmembership{Fellow,~IEEE}
        \thanks{The authors are with the School
of Electrical Engineering, Korean Advanced Institute of Science and Technology, Daejeon 34141, Korea.}
}
\maketitle

\begin{abstract}
The problem of radar detection in compound Gaussian clutter when
a radar signature is not completely known has not been considered yet and
is addressed in this paper.
We proposed a robust technique
to detect, based on the generalized likelihood ratio test, a point-like target 
embedded in compound Gaussian clutter.
Employing an array of antennas, we assume that
the actual steering vector departs from 
the nominal one, but lies in a known interval.
The detection is then secured by employing a semi-deﬁnite
programming.
It is confirmed via simulation that the proposed detector
experiences a negligible detection
loss compared to an adaptive normalized matched filter in a perfectly matched case,
 but outperforms in cases of mismatched signal.
 Remarkably, the proposed detector possesses constant false alarm rate with
respect to the clutter covariance matrix.
\end{abstract}


\IEEEpeerreviewmaketitle
\section{Introduction}
\IEEEPARstart{T}{he} problem of radar detection in Gaussian clutter
has been addressed in the
pioneering work \cite{Kelly86}. Therein, the
presence of a point-like target was sought in a single vector
of the form $b \boldsymbol{s}$, where
$b$ was an unknown complex scalar accounting
for the combined effect of  a target's reflectivity and channel propagation and
$\boldsymbol{s}$, representing the radar signature, was perfectly known.
It was reported that
the detector in \cite{Kelly86} suffers a
performance loss when
the actual radar signature
departs from its nominal one, for example cases of
an imperfect array calibration.

 To increase detection probability when a mismatch occurs,
 detecting $b \boldsymbol{s}$ where $\boldsymbol{s}$ is not completely known but lies
 in an assumed range has been proposed.
Such a range where $\boldsymbol{s}$ lies in could be possibly
modelled as a known linear subspace or a cone with axis the nominal radar signature.
Subspace detectors, based on the former approach, have been proposed in
 \cite{Scharf94}--\cite{Aubry_Jul14}.
 Coordinates of the signal to detect 
 are unknown; detection is performed
    by computing energy
of the measurement in the signal subspace \cite{Scharf94}.
However, there is no guidance on choosing an appropriate subspace to which
a signal of interest belongs.
The latter approach circumvents this drawback by
assuming a nominal radar signature as axis of a cone to which
a signal of interest belongs to \cite{Ramprashad96}.
Cone class based detectors
have been proposed in \cite{DeMaio05}--\cite{DeMaio09}:
In most cases, likelihood ratios
are obtained by numerical
methods, hence
it is difficult
to explain and investigate the detection
nature and performance.

 In the above mentioned research, the radar clutter is modelled as a
 Gaussian process, whereas
 in many circumstances, for instance under a low aspect angle,
  radar clutter is
   better characterized as a
 spherically invariant random process (SIRP)
 (compound Gaussian process) \cite{Conte87}\cite{Ward90}.
 Briefly, SIRP is
   a Gaussian process $g(t)$ (called \textit{speckle})
modulated by a temporally and spatially "more-slowly varying"
non negative random process $s(t)$ (called \textit{texture}), 
which is independent of $g(t)$ and represents the
 illumination patch's reflectivity.
 Problems of detecting a perfectly known 
 radar signature in compound Gaussian clutter
have been addressed in
 \cite{Conte95}--\cite{Rabaste_Jul14}, where detectors are called
 normalized matched filters.
 The problem of detecting in compound Gaussian clutter
  a mismatched signal, a possibility in
 some practical cases, has been not considered yet and
 is solved 
 in this paper.

 We addressed, based on the generalized likelihood ratio test (GLRT),
 the problem of detecting
 a point-like target 
 embedded in compound Gaussian clutter,
  employing
 an uniform array of antennas.
 Here, the radar signature is the steering vector that, due to some reason,
 departs from the nominal one.
 The maximum likelihood estimate (MLE) of the
 unknown steering vector lying in a cone 
 then leads to
  a fractional quadratically constrained quadratic optimization problem,
 which is not easy to solve \cite{DeMaio11}.
We hence introduced a more specific constraint on the mismatched steering vector:
phase shifting 
 of the mismatched steering vector lying in a known range.
 A practical example demonstrating the rationale of this assumption is
 a case of inaccurate estimate of an arrival angle.
 The optimization problem associated to the mismatched steering vector estimate
then can be transferred in a form
solvable via a semi--definite programming (SDP) \cite{Roh06}.
In case of a perfect match, 
the proposed detector provides a comparable detection probability with
that of a normalized matched filter. 
In the presence of a mismatch, 
the proposed detector outperforms a normalized matched filter
even with a slight
mismatch. Additionally, 
with a maximum mismatch lying in the range designed 
numerical results showed a
detection loss of around $3$ dB.
Factors that affect the proposed detector's performance have been investigated.
Remarkably, the proposed detector
possess CFAR w.r.t the clutter covariance matrix.\\
\indent The rest of the manuscript is organized as follows.
The problem formulation is stated and the proposed detector is derived in Section II.
Numerical results are represented in section III. Finally, conclusion is reported in section IV. \\
\indent \textit{Notation:}
We adopt the notation of using boldface lower case and upper case for vectors
and matrices, respectively. The transpose and complex conjugate transpose of a matrix
are denoted by $(\cdot)^T$ and $(\cdot)^H$, respectively.
For a square matrix,
$\tr(\cdot)$, $|\cdot|$, and $\rank(\cdot)$ respectively stand for its trace,
determinant, and rank.
$\diag(\boldsymbol{A})$ denotes a vector whose $i$-th element is the
$i$-th diagonal element of a matrix $\boldsymbol{A}$; while
$\diag(\boldsymbol{a})$ denotes
 a diagonal matrix whose diagonal elements are elements of a vector $\boldsymbol{a}$.
  $(\cdot)^{-1}$ represents the inverse of an invertible square matrix.
  $\circ$ denotes the Hadamard product.
  $\boldsymbol{A}_{nm}$ denotes the element at
  $n$-th row, $m$-th column of a matrix $\boldsymbol{A}$.
  $\mathbb{C}^{N \times N}$, $\mathbb{R}^{N \times N}$, and $\mathbb{H}^{N \times N}$
stand for the set of $N \times N$ complex, $N \times N$ real, and $N \times N$
Hermittian matrices, respectively.
For any $\boldsymbol{A} \in \mathbb{H}^{N \times N}$,
$ \boldsymbol{A} \succeq 0$ means that $\boldsymbol{A}$
is a positive semi-definite matrix.
$\mathbb{R}^{+}$ is the set of non-negative real numbers.
The real part of a complex scalar, vector, or matrix is represented by
$\Re \{\cdot \}$.
$\| \cdot \|$ is the Euclidean norm of a vector, and
$|\cdot|$ denotes the modulus of a complex number.
Finally, the letter $j$ represents the square root of $-1$ and $E[\cdot]$ denotes a statistical expectation.

\section{Problem Formulation and Proposed Detector}

\subsection{Problem Formulation}
Consider a problem of detecting the presence of a point-like target
using an uniform linear array of N antennas.
For the detection, reflection 
is collected at the cell under test (CUT) and surrounding range cells, in which
data is assumed to compose of only noise
and referred to as secondary data.
Target's return
 in an equivalent baseband form
is represented as
$\alpha \boldsymbol{p}$, where $\alpha$ is a complex scalar accounting for the combined effect of
a target's reflection and channel propagation
and $\boldsymbol{p}$ is an $N \times 1$ steering vector departing
from the nominal steering vector $\boldsymbol{s}$,
i.e.,
$\boldsymbol{s} = \left[1,\exp(j\theta),\ldots,\exp(j(N-1)\theta)\right]^T $
and
$\boldsymbol{p} = \left[1,\exp(j\phi),\ldots,\exp(j(N-1)\phi)\right]^T $, 
where
$(\theta - \phi)$ is unknown but 
 $(\theta - \phi) \in [-\beta,\beta]$ with $\beta$
  is a known quantity expressing
  the discrepancy of $\boldsymbol{p}$ from $\boldsymbol{s}$.
Clutter at a range cell $\boldsymbol{c}_t$ is modelled as
a complex SIRP
with index $ t = 0$ indicating the CUT and
$t = 1,2,\ldots,K$ indicating surrounding range cells.
From the definition of a complex SIRP \cite{Conte87}, we have
\begin{equation}\label{noise_model}
      \boldsymbol{c}_t \thinspace=\thinspace s_t \boldsymbol{g}_t, \qquad t = 0,1,2,\ldots, K.
\end{equation}
where $s_t$, the
\textit{texture} component of $\boldsymbol{c}_t$,
is a real non-negative random variable with some 
distribution $f_s(s_t)$ and
$\boldsymbol{g}_t$, the
\textit{speckle} component of $\boldsymbol{c}_t$,
is an $N \times 1$ vector of zero mean and multivariate complex normal distribution
with normalized covariance matrix $\boldsymbol{M}$, i.e.,
$\tr{\boldsymbol{M}} = N$.
We also assume that $\left \{ \boldsymbol{g}_0,\boldsymbol{g}_1,\ldots,\boldsymbol{g}_K  \right \}$
 are independent identically distributed
 and possess the circular symmetric property.
Note that $s_t$ and $\boldsymbol{g}_t$ are independent, 
$\left \{ s_t\right\}_{t = 0}^{t = K}$
may be correlated.
In practice, the distribution $f_s(s_t)$ is \textit{priori} unknown,
making it impossible to derive the probability density function (pdf)
of $\boldsymbol{c}_t$.
However,
if the illumination time is much shorter than
the de-correlation time of the \textit{texture} $s_t$
we can consider $s_t$ as, as in this paper,
unknown deterministic parameters \cite{Conte_Aug02}.
 This assumption then leads to the independence of clutter
 at all range cells.

The detection problem can now be stated as a problem of binary hypotheses
\begin{equation}\label{hypotheses}
 \left \{
 \begin{array}{ll}
   H_0 : & \boldsymbol{z} \thinspace=\thinspace
\boldsymbol{c}_0,  \\
   H_1 : &  \boldsymbol{z} \thinspace=\thinspace
\alpha \boldsymbol{p} + \boldsymbol{c}_0,
 \end{array}
  \right.\
\end{equation}
where the null hypothesis $H_0$ and alternative hypothesis $H_1$ denote the cases of
clutter-only and signal plus clutter, respectively, and $\boldsymbol{z}$ denotes the equivalent baseband of received signal at the CUT.
The pdf of the observed data $\boldsymbol{z}$
can be expressed as
\begin{equation}\label{null_hp}
f_0\left(\boldsymbol{z}
 \right) \thinspace=\thinspace
 \frac{1}{\pi^N s_0^{2N}|\boldsymbol{C}|}
 \exp \left \{ -\boldsymbol{z}^H \frac{1}{s_0^2} \boldsymbol{C}^{-1}\boldsymbol{z} \right \}
\end{equation}
and
\begin{multline}\label{alt_hp}
f_1\left(\boldsymbol{z}
 \right) \thinspace=\thinspace \\
 \frac{1}{\pi^N s_0^{2N}|\boldsymbol{C}|}
 \exp \left \{ -\left(\boldsymbol{z} - \alpha \boldsymbol{p}\right)^H \frac{1}{s_0^2}
 \boldsymbol{C}^{-1}\left(\boldsymbol{z} - \alpha \boldsymbol{p}\right) \right \},
\end{multline}
where $\boldsymbol{C} = E[\boldsymbol{c}_t \boldsymbol{c}_t^H]$ the covariance matrix of radar clutter.
It's easy to see that $\boldsymbol{C} = E[s^2]\times \boldsymbol{M}$.
To maximise the detection probability
given a predetermined false alarm rate,
we employ the Neyman-Pearson criterion.
Due to the ignorance of
the clutter covariance matrix $\boldsymbol{C}$, \textit{texture} $s_0$, the steering vector
$\boldsymbol{p}$, and $\alpha$, we resort to a GLRT scheme,
replacing these nuisance parameters with their MLEs under each hypothesis
\begin{equation}\label{glrt}
  \frac{\max \limits_{\boldsymbol{p}} \max \limits_{\alpha} \max \limits_{s_0} \max \limits_{\boldsymbol{C}}
   f_1\left(\boldsymbol{z} \right)}
  {\max \limits_{s_0} \max \limits_{\boldsymbol{C}} f_0\left(\boldsymbol{z}\right)}
  \thinspace\mathop{\gtrless}_{H_0}^{H_1}\thinspace G_2,
\end{equation}
where $G_2$ is a threshold set for a predetermined false alarm rate.
For detection, the next logical step is to replace $\boldsymbol{C}$ with its MLE.
However, it is proved in \cite{Gini99} that
a closed-form of the
MLE of covariance matrix $\boldsymbol{C}$
does not exist.
Hence, we
assumed known $\boldsymbol{C}$
in the following development
 and derive
a detector.
Later, $\boldsymbol{C}$ is replaced by some estimators
and properties of the resulting detectors
will be discussed. 
\subsection{Detection with known structure of covariance matrix $\boldsymbol{C}$}
MLEs of the \textit{texture} components under each hypothesis are given as in \cite{Conte_Apr02}
\begin{eqnarray}\label{texture_mle}
  H_0 \thinspace : \thinspace s_0^2 &=& \frac{1}{N} \boldsymbol{z}^H \boldsymbol{C}^{-1} \boldsymbol{z}, \\
  H_1 \thinspace : \thinspace s_0^2 &=& \frac{1}{N}
\left ( \boldsymbol{z} - \alpha \boldsymbol{p} \right  )^H \boldsymbol{C}^{-1}\left ( \boldsymbol{z} - \alpha \boldsymbol{p} \right  ).
\end{eqnarray}
Direct substitution of the MLEs
of $s_0$ into (\ref{glrt}) leads to
\begin{equation}\label{glrt_alphaCp}
\frac{\left(\boldsymbol{z}^H \boldsymbol{C}^{-1} \boldsymbol{z}\right)^N}
{\min \limits_{\alpha}
 \min \limits_{\boldsymbol{p}}
  \left \{ \left ( \boldsymbol{z} - \alpha \boldsymbol{p} \right  )^H \boldsymbol{C}^{-1}\left ( \boldsymbol{z} - \alpha \boldsymbol{p} \right  )\right\}^N
  }
 \thinspace\mathop{\gtrless}_{H_0}^{H_1}\thinspace G_1,
\end{equation}
where $G_1$ is a suitable modification of $G_2$.
We then proceed by replacing $\alpha$ with its MLE, which is \cite{Conte95}
\begin{equation}\label{alpha_mle}
\alpha \thinspace = \thinspace
\frac{\boldsymbol{p}^H \boldsymbol{C}^{-1} \boldsymbol{z}}
{\boldsymbol{p}^H \boldsymbol{C}^{-1} \boldsymbol{p}}.
\end{equation}
into (\ref{glrt_alphaCp}).
After some manipulations, the likelihood ratio is recast as
\begin{equation}\label{glrt_equi}
\max \limits_{\boldsymbol{p}}
\frac{\left |\boldsymbol{z}^H \boldsymbol{C}^{-1}\boldsymbol{p} \right |^2}
     { \left (\boldsymbol{z}^H \boldsymbol{C}^{-1} \boldsymbol{z} \right )
   \left (\boldsymbol{p}^H \boldsymbol{C}^{-1} \boldsymbol{p} \right )}
   \thinspace\mathop{\gtrless}_{H_0}^{H_1}\thinspace G,
\end{equation}
where $G$ is a suitable modification of $G_1$.
It is easy to see,
by using the Schwatz's inequality,
 that $G \in [0,1]$.
Also, the test in (\ref{glrt_equi}) does not change if
we substitute $\boldsymbol{C}$ with $\boldsymbol{M}$.
In addition,
if $\boldsymbol{p}$ is known,
(\ref{glrt_equi}) becomes the likelihood ratio
of the detector proposed
in \cite{Conte95},
\begin{equation}\label{asymp_lrt}
\frac{\left |\boldsymbol{z}^H \boldsymbol{M}^{-1}\boldsymbol{p} \right |^2}
     { \left (\boldsymbol{z}^H \boldsymbol{M}^{-1} \boldsymbol{z} \right )
   \left (\boldsymbol{p}^H \boldsymbol{M}^{-1} \boldsymbol{p} \right )}
   \thinspace\mathop{\gtrless}_{H_0}^{H_1}\thinspace G.
\end{equation}
The detector in \cite{Conte95}
(\ref{asymp_lrt}) is referred to as the normalized matched filter (NMF)
with known $\boldsymbol{M}$
and as an adaptive NMF (ANMF) with an estimated $\boldsymbol{M}$.
It is worth noting that (\ref{asymp_lrt}) was derived in \cite{Conte95}
to detect a coherent pulse trains with 
the number of pulses goes to infinity, given that
$s_0$ was a
random variable of a well-behaved distribution.

\indent
Now, we have to solve the maximization problem in (\ref{glrt_equi})
w.r.t $\boldsymbol{p}$ before proceeding to a decision on a target's presence.
From the observation that the expression to be maximized
depends only on $\phi$,
we rewrite (\ref{glrt_equi}) as
\begin{equation}\label{glrt_Fphi}
\frac{1}{\boldsymbol{z}^H \boldsymbol{C}^{-1} \boldsymbol{z}}
\left [
\max \limits_{\phi \in [\theta - \beta,\theta + \beta]}F(\phi)
\right ]
   \thinspace\mathop{\gtrless}_{H_0}^{H_1}\thinspace G_2,
\end{equation}
where
$
F(\phi) \thinspace = \thinspace
                          \dfrac{\boldsymbol{p}^H \boldsymbol{C}^{-1}\boldsymbol{z}\boldsymbol{z}^H \boldsymbol{C}^{-1}\boldsymbol{p}}
                          {\boldsymbol{p}^H \boldsymbol{C}^{-1} \boldsymbol{p}}
$.
Note that $F(\phi)$ is always non-negative since
$\boldsymbol{C}^{-1}$
and
$\boldsymbol{C}^{-1}\boldsymbol{z}\boldsymbol{z}^H \boldsymbol{C}^{-1}$
 are semi-definite.
Expanding
$F(\phi)$ in term of $\phi$ and using
$\exp (-j k \phi) = \left (\cos{\phi} - j \sin{\phi}\right )^k$, we have
\begin{equation}\label{Fphi_intheta}
F(\phi) \thinspace = \thinspace  \frac
           {x_0 + 2 \Re \sum \limits_{k=1}^{N-1} x_k \exp(-jk\phi) }
           {y_0 + 2\Re \sum \limits_{k=1}^{N-1} y_k \exp(-jk\phi) }
\end{equation}
with
\begin{eqnarray}
    x_0 \thinspace &=& \thinspace \tr \left(\boldsymbol{C}^{-1}\boldsymbol{z}\boldsymbol{z}^{H}\boldsymbol{C}^{-1}\right), \label{x0_poly} \\
 y_0 \thinspace &=& \thinspace \tr \left(\boldsymbol{C}^{-1}\right), \\
  x_k \thinspace &=& \thinspace \sum\limits_{m-n=-k} \left(\boldsymbol{C}^{-1}\boldsymbol{z}\boldsymbol{z}^{H}\boldsymbol{C}^{-1}\right)_{nm}, \\
  y_k \thinspace &=& \thinspace \sum\limits_{m-n=-k} \left (\boldsymbol{C}^{-1}\right)_{nm},\label{yk_poly}
\end{eqnarray}
and $k= 1,2,\ldots,N-1$.
 Note that $x_0$ and $y_0$ are real since
$ \boldsymbol{C}^{-1}\boldsymbol{z}\boldsymbol{z}^{H}\boldsymbol{C}^{-1} $ and
$\boldsymbol{C}^{-1}$
are Hermitian.
We observe that
finding the maximum w.r.t $\phi$ of $F(\phi)$ in (\ref{Fphi_intheta}) is not straightforward
since the numerator and denominator are polynomials of
at most ($N-1$)-th degree in $\cos{\phi}$ and $\sin{\phi}$.
We then solve the maximization here by a numerical method.
Firstly, denote by $t$ the maximum value of $F(\phi)$, then $t$ is the
lowest upper bound of $F(\phi)$, i.e. $t$ is the solution of
the optimization problem
 \begin{equation}\label{Fphi_minimization}
\begin{array}{ll}
 \underset{t \in \mathbb{R}^{+}}{\text{minimize}} & t \\
                           \text{such that} &
                           g(\phi, t)\thinspace \triangleq \thinspace \geqslant 0, \\
 & \phi \in [\theta - \beta,\theta + \beta].
  \end{array}
\end{equation}
where
 $$
  g(\phi, t)\thinspace \triangleq \thinspace ty_0 - x_0 +
                           2\Re \left\{ \sum_{k=1}^{N-1}\left(ty_k - x_k\right)\exp(-jk\phi)\right\}
 $$

We have another observation that
$g(\phi, t)$ is
a real non-negative trigonometric polynomial
over the interval
$[\theta - \beta,\theta + \beta]$, so
coefficients of $g(\phi, t)$ follow the following theorem \cite{Davidson02}.
\begin{definition}
Let $
\boldsymbol{W}_{DFT} \in \mathbb{C}^{M \times M}
$
be the DFT matrix
\begin{equation*}
  \boldsymbol{W}_{DFT} = \left [ \boldsymbol{w}_0, \boldsymbol{w}_1,\ldots,\boldsymbol{w}_{M-1} \right ],
\end{equation*}
where
$
\boldsymbol{w}_k = \left[ 1, \exp(-jk2\pi/M),\ldots,\exp(-j(M-
\right. \linebreak \left.
1)k 2\pi/M)\right]^T
$.
\end{definition}
We define $\boldsymbol{W}$ and $\boldsymbol{W}_1$ as matrices composed of the first $N$ and $N-1$
columns of $\boldsymbol{W}_{DFT}$, respectively.
\begin{theorem}
Let $p(\phi)$ be a trigonometric polynomial in $\phi$ with degree (N-1) or less, and have the form
$$
p(\phi)\thinspace = \thinspace q_0 + 2\Re \sum \limits_{k =1}^{N-1} q_k \exp \left(-jk\phi \right),
 $$
 with $\boldsymbol{q} = [q_0, q_1,q_2,\ldots,q_{N-1}] \in \mathbb{R} \times \mathbb{C}^{N-1}$.
 $p(\phi)$ is non-negative on $[\theta - \beta,\theta + \beta]$ 
 if and only if there exist
 $\boldsymbol{X}_1 \in \mathbb{H}^{N \times N}$ and $\boldsymbol{X}_2 \in \mathbb{H}^{(N-1)\times (N-1)}$ so that
 $ \boldsymbol{q} =
 \boldsymbol{W}^H \left [\diag \left ( \boldsymbol{W} \boldsymbol{X}_1 \boldsymbol{W}^H \right ) +
 \boldsymbol{d}\circ \diag \left ( \boldsymbol{W}_1 \boldsymbol{X}_2 \boldsymbol{W}_1^H \right )\right]$,
 where
 $\boldsymbol{X}_1 \succeq 0$, $\boldsymbol{X}_2 \succeq 0$,
  $\boldsymbol{d} \in \mathbb{R}^{M \times 1}$ has elements
 $d_k = \cos(2\pi k/M-\theta)-\cos\beta$ for $k = 0,1,\cdots, M-1$
 and $M \geq 2N - 1$.
 \end{theorem}
 For a detailed explanation of the idea underlying the above theorem as well as its applications,
 interest readers may refer \cite{Roh06}.
 Applying the theorem and
 denoting
 $\boldsymbol{y} = \left [y_0, y_1,\ldots,y_{N-1}\right ]^T$ and
 $\boldsymbol{x} = \left [x_0, x_1,\ldots,x_{N-1}\right]^T$, with
 $x_i$, $y_i$, $ i = 0,1,\ldots,N-1$ are computed as in
 (\ref{x0_poly})--(\ref{yk_poly}),
  we have
  $
   t \boldsymbol{y} - \boldsymbol{x}
   \thinspace  = \thinspace
   \boldsymbol{W}^H \left [\diag \left ( \boldsymbol{W} \boldsymbol{X}_1 \boldsymbol{W}^H \right ) + \\
 \boldsymbol{d}\circ \diag \left ( \boldsymbol{W}_1 \boldsymbol{X}_2 \boldsymbol{W}_1^H \right )\right],
 $

\indent The minimization (\ref{Fphi_minimization}) is now recast as a SDP
\begin{equation}\label{SDP_detector}
\begin{array}{ll}
                        \underset{t,\boldsymbol{X}_1,\boldsymbol{X}_2}{\text{minimize}} & t\\
 \text{s.t} & t\boldsymbol{y}-\boldsymbol{x} \thinspace=\thinspace
 \boldsymbol{W}^H [ \diag \left ( \boldsymbol{W} \boldsymbol{X}_1 \boldsymbol{W}^H \right ) + \\
            & \boldsymbol{d}\circ \diag \left ( \boldsymbol{W}_1 \boldsymbol{X}_2 \boldsymbol{W}_1^H \right ) \\
                            & t \in \mathbb{R}^{+} \\
                            & \boldsymbol{X}_1 \succeq 0,\thinspace \boldsymbol{X}_1 \in \mathbb{H}^{N \times N} \\
                              & \boldsymbol{X}_2 \succeq 0, \boldsymbol{X}_2 \in \mathbb{H}^{(N-1)\times (N-1)} ].
        \end{array}
\end{equation}
The SDP above can be solved efficiently using the interior point method.
In passing, we note that
 this algorithm was applied in \cite{DeMaio10} to  detect a point-like target
in correlated Gaussian noise under an unknown direction of arrival.\\
\indent Denote by $t^{\ast}$ the optimal value 
attained from solving (\ref{SDP_detector}),
the likelihood ratio  test is written as follows
\begin{equation}\label{final_test}
\frac{1}
     {  \boldsymbol{z}^H \boldsymbol{C}^{-1} \boldsymbol{z} } t^{\ast}
   \thinspace\mathop{\gtrless}_{H_0}^{H_1}\thinspace G_2.
\end{equation}
\indent

\subsection{Detection with estimated structure of covariance matrix $\boldsymbol{C}$}
As mentioned in the previous subsection,
the likelihood ratio (and its statistic) in (\ref{glrt_equi})
remains unchanged if we substitute
$\boldsymbol{C}$ by $\boldsymbol{M}$.
Hence, instead of estimating $\boldsymbol{C}$, we employ
an estimate of $\boldsymbol{M}$.
The MLE of $\boldsymbol{M}$ 
has been proved to uniquely exist and derived in
\cite{Pascal08}, in which
MLE of $\boldsymbol{M}$
satisfies the equation
\begin{equation}\label{MLE_compoundG}
     \boldsymbol{M}_{MLE} \thinspace = \thinspace f(\boldsymbol{M}_{MLE}),
\end{equation}
where 
\begin{equation}\label{MLE_func}
  f(\boldsymbol{M}_{MLE}) \thinspace = \thinspace
  \frac{N}{K} \sum_{t=1}^{K} \frac{ \boldsymbol{c}_t \boldsymbol{c}_t^H}
                                  {\boldsymbol{c}_t^H \boldsymbol{M}_{MLE}^{-1}\boldsymbol{c}_t}.
  \end{equation}
  Solution for the above equation uniquely exists but
a closed-form for such solution does not exist \cite{Pascal08}.
Instead, MLE of $\boldsymbol{M}$ is computed by recursion computing \cite{Pascal08},
which
is employed in this paper.

Now, we replace $\boldsymbol{C}$ in (\ref{glrt_equi}) by the MLE of $\boldsymbol{M}$
and assess the CFAR property of the resulting detector,
 called $\theta-$MLE detector, employing the likelihood ratio
\begin{equation}\label{proposed_lrt}
  \frac{1}
     {  \boldsymbol{z}^H \widehat{\boldsymbol{M}}^{-1} \boldsymbol{z} } \hat{t}^{\ast}
   \thinspace\mathop{\gtrless}_{H_0}^{H_1}\thinspace G,
\end{equation}
where
$\hat{t}^\ast$ is the optimal value attained from (\ref{SDP_detector}), in which
$\boldsymbol{y}$ and $\boldsymbol{x}$ are computed with
$\boldsymbol{M}_{MLE}$.
It is easily to see that the
$\theta-$MLE
detector have CFAR w.r.t \textit{texture} components
$\left\{s_0,s_1,s_2,\ldots,s_K\right\}$. This claim is easily proved based on the
following arguments.
Firstly, notice that
the MLE of $\boldsymbol{M}$ can be derived based on the relation 
\begin{equation}\label{equi_mle}
  \boldsymbol{M}_{MLE} \thinspace = \thinspace
  \frac{N}{K} \sum_{t=1}^{K} \frac{ \boldsymbol{g}_t \boldsymbol{g}_t^H}
                                  {\boldsymbol{g}_t^H \boldsymbol{M}_{MLE}^{-1}\boldsymbol{g}_t},
\end{equation}
 which
 is independent of $\left \{s_0,s_1,s_2,\ldots,s_K \right \}$.
 In addition, the \textit{texture} component $s_0$ embedded in $\boldsymbol{z}$
 has been cancelled out in the numerator and denominator of
   $
 \dfrac{\left | \boldsymbol{z}^H  \boldsymbol{M}_{MLE} ^{-1}\boldsymbol{p} \right |^2}
     { \left (\boldsymbol{z}^H \boldsymbol{M}_{MLE}^{-1} \boldsymbol{z} \right )
   \left (\boldsymbol{p}^H \boldsymbol{M}_{MLE}^{-1} \boldsymbol{p} \right )}
 $.
 Hence,
  the likelihood ratio (\ref{proposed_lrt}) is independent of $\left \{s_0,s_1,s_2,\ldots,s_K\right \}$.
 Regarding the CFAR property w.r.t $\boldsymbol{M}$,
 it is very hard to analyze
 the dependence of  the false alarm rate of $\theta$-MLE detector on $\boldsymbol{M}$,
 such dependence will be numerically analyzed. 

\section{Numerical Results}
In this section, via computer simulation we assess and compare performance
 of the $\theta$-MLE detector (\ref{proposed_lrt})
with that of the ANMF \cite{Conte95}, referred to as in the following as MLE-NMF since
the unknown $\boldsymbol{M}$ is replaced with its MLE.
\begin{equation}\label{anmf_recall}
\frac{\left |\boldsymbol{z}^H    \boldsymbol{M}_{MLE}^{-1}\boldsymbol{p} \right |^2}
     { \left (\boldsymbol{z}^H \boldsymbol{M}_{MLE}^{-1} \boldsymbol{z} \right )
   \left (\boldsymbol{p}^H \boldsymbol{M}_{MLE}^{-1} \boldsymbol{p} \right )}
   \thinspace\mathop{\gtrless}_{H_0}^{H_1}\thinspace G.
\end{equation}
For the simulation,
 we use an uniform linear array consisting of $N = 8$ antennas,
 assuming $\theta = \pi/3$ and $\beta = \pi/6$
 (i.e. $\phi \in [\pi/6, \pi/2]$) and $K = 32$.
 As to the clutter, we assume that $s_0,s_1,\ldots,s_K$
 follow the chi distribution, so $s_0^2,s_1^2,\ldots,s_K^2$ follow the chi-square distribution
 with degree of freedom $\nu = 3$ (\cite{Ward81}), i.e. $E[s_t^2] = 3$.
 The generation of $\boldsymbol{g}_t$
follows the guide in \cite{Rangaswamy95}. Briefly,
we firstly generate complex Gaussian random vectors $\boldsymbol{u}_t$
 of zero-mean and identity covariance matrix; next
 $\boldsymbol{g}_t = \boldsymbol{R} \boldsymbol{u}_t $,
where
 $\boldsymbol{R}$ is the Cholesky decomposition of $\boldsymbol{C}$, i.e,
 $\boldsymbol{R}\boldsymbol{R}^H = \boldsymbol{C}$
 where $\boldsymbol{C}_{nm} = \rho^{|n-m|}$
 and $\rho$ the correlation efficient.
 Clutter return at each range cell is $\boldsymbol{c}_t = s_t \boldsymbol{g}_t$.
 Since it is difficult to derive closed-forms of detection ($P_d$)
and false alarm ($P_{fa}$) probabilities, such quantities will be numerically analyzed
through independent $10^2 / P_{d}$ and $10^2 / P_{fa}$ Monte Carlo trials, respectively.
To lower the computational burden, we choose $P_{fa} = 10^{-3}$.
We use the software CVX (http://cvxr.com/) to solve the semi--deﬁnite problem
 (\ref{proposed_lrt}) on a computer equipped with a 3.4 GHz Intel processor.
Finally, the signal-to-noise ratio (SNR) is defined as
 \begin{equation}\label{SNR}
  \text{SNR} \thinspace = \thinspace
  \frac{| \alpha |^2 \| \boldsymbol{p} \|^2 }
        {N \times E[s^2]},
\end{equation}
which is $| \alpha |^2 \| \boldsymbol{p} \|^2 /(N \times \nu )$ in our simulation.
\subsection{Performance Assessment}
\begin{figure}
  \includegraphics[width=21pc]{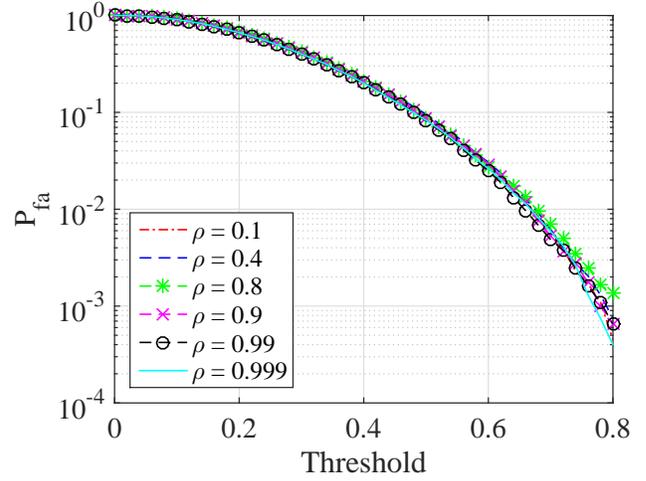}
  \caption{Probability of false alarm versus the detection threshold,
  $\theta$-MLE, $N = 8$, $K = 32$.}
  \label{pfa_thre_mle}
 \end{figure}

We first investigate if the $\theta$-MLE detector has a CFAR property
w.r.t structure of the clutter covariance matrix (i.e. $\boldsymbol{M}$).
Fig. \ref{pfa_thre_mle} shows false alarm probabilities versus threshold of
the $\theta$-MLE detector at varied degrees of correlation
$\rho = 0.1, 0.4, 0.8, 0.9, 0.99, 0.999$.
Here we used $5 \times 10^4$ Monte Carlo runs.
It is observed that $\theta$-MLE possesses CFAR w.r.t all simulated degrees of correlation.
Hence, $\theta$-MLE detector possesses CFAR w.r.t all the statistics of the clutter,
a property that is also possessed by MLE-NMF \cite{Conte_Aug02}.
From now on, $\rho = 0.4$ in all simulations.

 \begin{figure}
  \includegraphics[width=21pc]{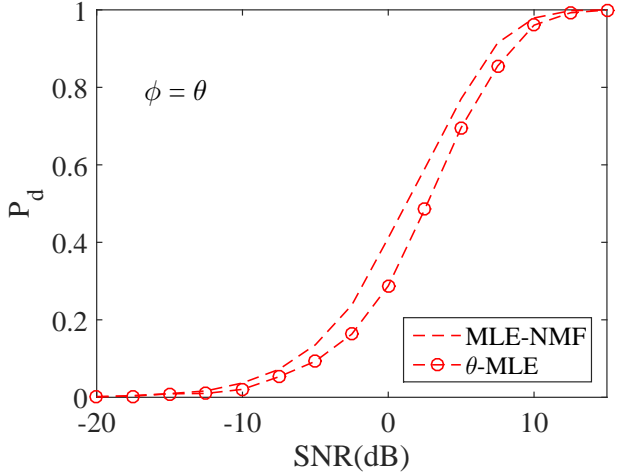}
  \caption{Detection probabilities versus SNR of $\theta$-MLE in comparison with MLE-NMF
   in perfectly matched case ($\phi = \theta$),
  $N = 8$, $K = 32$.}
  \label{pd_matched_thetaVSnmf}
 \end{figure}

  \begin{figure}
  \includegraphics[width=21pc]{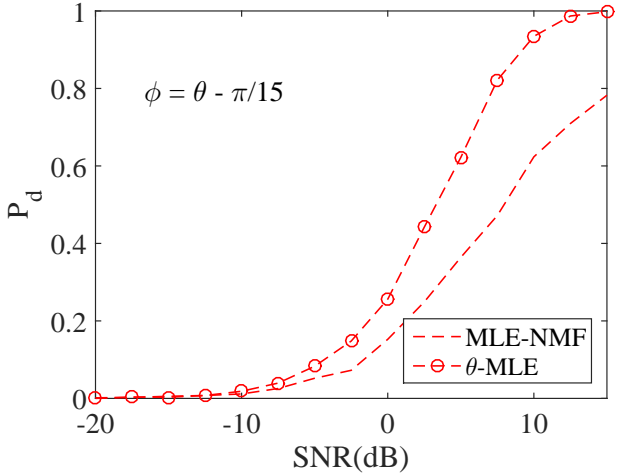}
  \caption{Detection probabilities versus SNR of $\theta$-MLE in comparison with MLE-NMF
  in mismatched case ($\phi = \theta - \pi/15$), $N = 8$, $K = 32$.}
  \label{chi_pd_mismatched_thetaVSnmf}
 \end{figure}

  \begin{figure}
  \includegraphics[width=21pc]{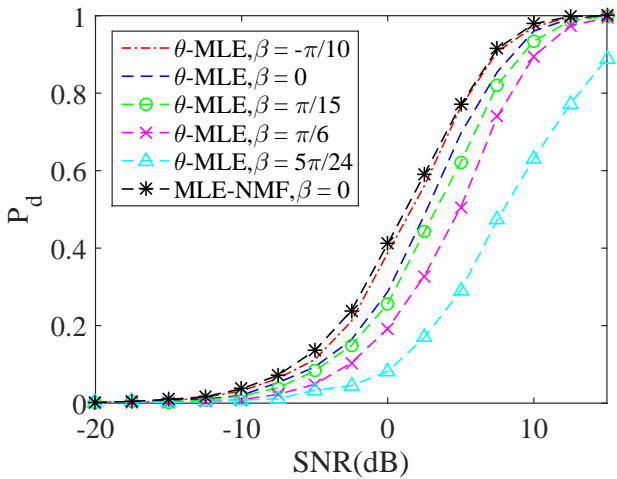}
  \caption{Detection probabilities versus SNR
  of $\theta$-MLE in varied values of $\beta$,
  $N = 8$, $K = 32$.}
  \label{chi_pd_mismatched_mle_theta_variedBeta}
   \end{figure}

  \begin{figure}
  \includegraphics[width=21pc]{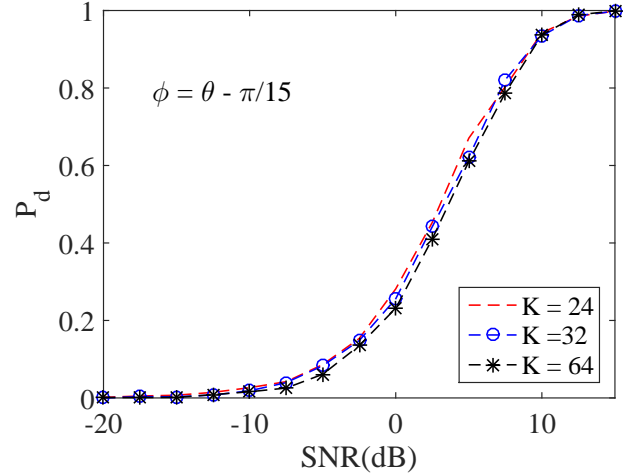}
  \caption{Detection probabilities versus SNR of $\theta$-MLE
  in varied values of K,
  $\beta = \pi/15$, $N = 8$.}
  \label{chi_pd_mismatched_scmAndmle_theta_variedK}
 \end{figure}

\indent In Fig. \ref{pd_matched_thetaVSnmf} we compare
detection probabilities of the $\theta$-MLE
 with that of the
MLE-NMF in case that the actual steering vector $\boldsymbol{p}$
perfectly matches with the nominal one $\boldsymbol{s}$,
i.e. $\phi = \theta$.
With incomplete knowledge of the actual steering vector,
$\theta$-MLE suffers a detection loss w.r.t
that of the MLE-NMF,
defined as the horizontal displacement of the corresponding curves,
of nearly 2dB. However, it is obvious and
shown in Fig. \ref{chi_pd_mismatched_thetaVSnmf} that even with a slight mismatch,
 i.e., $\theta - \phi= \pi/15$,
  $\theta$-MLE
  outperforms the
  MLE-NMF, especially in the high SNR region.
  Robustness of the $\theta$-MLE detector to mismatched signal is further demonstrated, in
  Fig. \ref{chi_pd_mismatched_mle_theta_variedBeta}
  in cases of more serious mismatches, i.e., $\theta - \phi = -\pi/10, 0, \pi/15, \pi/6, \pi/24$.
  Loss in detection probabilities, in a comparison with perfectly matched case,
   of MLE-NMF
   is comparatively small
   when a mismatch lies in the designed interval of the $\theta$-MLE
    and becomes significant with a mismatch lying outside
 the designed interval, i.e., $|\theta - \phi| > \beta$, i.e in case of
 $\theta - \phi = 5\pi/24$ (the designed $\beta = \pi/6$).
    Finally, influence on $\theta$-MLE's detection performance of the size of secondary data
 is investigated in Fig. \ref{chi_pd_mismatched_scmAndmle_theta_variedK}.
Interestingly, $P_d$ of the proposed detector exhibits a little improvement with an increasing value of $K$,
meaning that we do not need to collect more secondary data from the surrounding range cells to
achieve better detection capacity.
This property is also
reported in the previous research \cite{Conte95}--\cite{Conte_Aug02} and is opposite to the results in
case of homogeneous/partially homogeneous Gaussian noise \cite{Conte01}.

\section{Conclusion}
This paper has addressed the problem of detecting a mismatched signal embedded in compound Gaussian noise.
Specifically, phase shifting of the actual steering vector departs from that of the nominal one but
belongs to a known interval.
The proposed detector is shown to be more robust to mismatched signals than the adaptive NMF,
and even achieves reasonable detection probabilities when the signal to detect lying out of the designed interval.
 Remarkably, the $\theta$-MLE detector has CFAR w.r.t all statistic of noise.
 A drawback of the proposed detector is that the likelihood ratio has no explicit form, for which
 it is difficult to gain a deeper insight into the performance of the detector.
Another drawback is the complexity associated with the SDP.
Though proposed scheme can detect a seriously mismatched signal, it does not include
effects of possible interference, which might be a topic for a further research.

\section*{Acknowledgment}
The authors would like to thank the anonymous reviewers for their valuable comments and suggestions to improve the paper.
\ifCLASSOPTIONcaptionsoff
  \newpage
\fi
 
%





\end{document}